# QoS Management Mechanisms for Enhanced Living Environments in IoT

Yassine Banouar[1, 2], Clovis Anicet Ouedraogo[1], Christophe Chassot[1, 3], Abdellah Zyane[4]

[1] CNRS, LAAS, 7 avenue du Colonel Roche, F-31400 Toulouse, France
Univ. de Toulouse, [2] UPS, [3] INSA, LAAS, F-31400 Toulouse, France
[4] Cadi Ayyad University, S.A.R.S Group, ENSA Safi, Morocco
{banouar, ouedraogo, chassot}@laas.fr, a.zyane@uca.ma

*Abstract*—The Internet of Things (IoT) paradigm is expected to bring ubiquitous intelligence through new applications in order to enhance living and other environments. Several research and standardization studies are now focused on the Middleware level of the underlying communication system. For this level, several challenges need to be considered, among them the Quality of Service (QoS) issue. The Autonomic Computing paradigm is now recognized as a promising approach to help communication and other systems to self-adapt when the context is changing. With the aim to promote the vision of an autonomic Middleware-level QoS management for IoT-based systems, this paper proposes a set of QoS-oriented mechanisms that can be dynamically executed at the Middleware level to correct QoS degradation. The benefits of the proposed mechanisms are also illustrated for a concrete case of Enhanced Living Environment.

## I.  Introduction

The Internet of Things (IoT) is expected to bring a large and promising spectrum of ubiquitous intelligence through new applications in various domains such as health, transport, industry, energy, environment, retail or logistics. Living environments should strongly benefit from those new communication capabilities that will make possible the execution of dynamic and autonomous actions going from, for instance in healthcare context, accurate supervision of patients to intervention in case of emergency. To make possible the deployment of such Enhanced Living Environments (ELE), network-level communication capabilities (such as Bluetooth, Zigbee, Wi-Fi, or wired technologies) are not sufficient. Indeed, Middleware-level platforms are required to allow hiding complexity of heterogeneous devices (sensors, effectors, tags, etc.) and communication technologies. Without such an intermediate abstraction layer between the IoT applications and the underlying communication technologies, the implementation of each application would be dependent on the devices technologies, leading to a vertical fragmentation between IoT applications and an impossible extensibility when new devices/capabilities appear.

In 2012, the ETSI standardization effort resulted in the proposition of the Machine-to-Machine (M2M) architectural framework [1] enabling the management of IoT applications, the abstraction of networks and devices heterogeneity, etc. This specification has been extended recently under the oneM2M architectural framework [2] which is now recognized as the de facto standard for the IoT paradigm.

Basically, both ETSI M2M and oneM2M frameworks are built on 4 levels: (1) *Application level* where business applications (home automation, smart metering, smart grid, etc.) offering final solutions are defined; (2) *Middleware level* that aims at hiding the details of various underlying networks and technologies to facilitate interoperability; (3) *Network level* that includes different types of networks to interconnect equipment; and (4) *Device level* that includes IoT sensor / actuators devices. An instantiation of this framework (Fig. 1) is composed of Business applications, Server and Gateways acting as Middleware entities, and finally Devices interconnected to their Gateways through specific network technologies (Bluetooth, Zigbee, etc.).

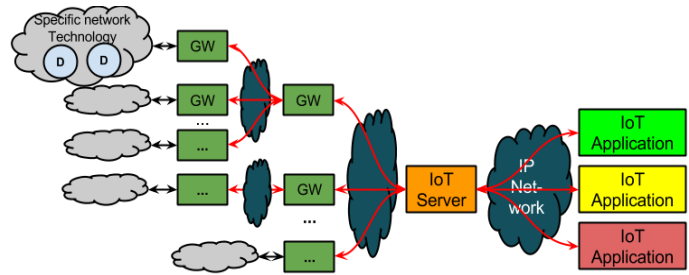

Fig. 1.  Instantiation of the oneM2M framework for IoT.

ELE are expected to be driven by several kinds of applications. In healthcare context, IoT can bring remote supervision of patients' health (heartbeat rate, glycaemia, etc.) and intervention when critical situation occurs. Each activity may be supported thanks to several devices held by the patients themselves, but also by all the entities involved in the global process (pharmacy, doctor, hospitals, etc.). All the exchanges have different priorities depending on the context: for instance, an alarm indicating a heart attack of a "risky-condition" patient is more important than supervision information of a young jogger. In terms of communication, such differences have to be translated into quality of service (QoS) requirements: the transmission delay of an alarm is expected to be as fast as possible, while a great delay of a non-sensitive supervision data could be acceptable.

Providing IoT platforms with QoS-oriented capabilities becomes a necessity that remains under research study [3, 4]. QoS issue in such environments is more considered at the network and device levels by the proposition of packet scheduling algorithms [5], service differentiation techniques [6], routing protocols [7] and adaptive architectures for devices [8]. The work presented in this paper focuses on the Middleware level and aims at promoting the vision of a dynamic and autonomic management of QoS-oriented adaptive actions, following the autonomic computing paradigm (AC) [9]. As depicted in Fig. 2, this paradigm is based on the monitoring, analysis, planning and execution steps, which are aimed at applying adaptation actions on the managed system according to its state thanks to a shared knowledge base. In our case, the considered system is an ETSI-M2M compliant Middleware platform, the open-source OM2M platform [10].

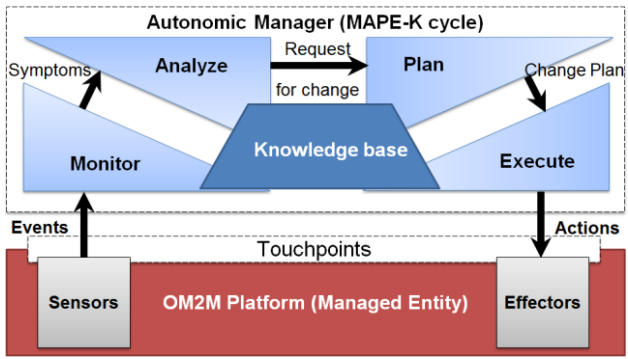

Fig. 2. Deployment context of the proposed autonomic computing-based approach for Middleware-level QoS management.

The contributions of this paper are focussed on the adaptation actions that may be performed to go toward the respect of the required QoS. Two kinds of adaptation actions are considered: "request-oriented" adaptation deals with the differentiation actions that can be done in the processing of requests having different QoS requirements; "resource-oriented" actions deals with the adaptation of the deployment resources, for instance at the virtual machine level for Middleware-level entities that are deployed in a cloud environment (typically IoT Servers).

The rest of this paper is structured as follows. Section II describes the QoS-oriented mechanisms that we propose at the Middleware level for an ETSI M2M compliant IoT platform. Section III illustrates how some of those mechanisms can be implemented within a typical example of ELE scenario. Performance measurements allow showing the benefits that can be induced with the application of those mechanisms. Finally, section V provides conclusion and future work.

## II. QOS-MANAGEMENT MECHANISMS AT THE MIDDLEWARE LEVEL

In order to satisfy the requirements of critical IoT applications, Middleware-level functionalities need to be extended to include QoS management mechanisms. In this section, we propose two levels of mechanisms: (1) *request-oriented* mechanisms that intervene on incoming HTTP requests; and (2) *resource-oriented* mechanisms that re-configure Middleware resources. Those mechanisms are supposed to be configured by an autonomic manager (AM), and more specifically by its planning component, according to the QoS constraints of critical IoT applications but also to the type of the managed entity (Server or Gateway) and its current performance state.

### A. Request-oriented Mechanisms

Request-oriented mechanisms are inspired by the mechanisms used in the Internet, specifically at the Transport and Network layers of the TCP/IP architecture. The approach follows the main principles of the DiffServ model [11], which is based on packet marking at the network entry and differentiation of the packet processing in each core routers depending on their mark. To implement a similar approach at the Middleware level, we add a field to the HTTP request specifying its priority; this priority allows guiding the request processing through the mechanisms components.

Two successive steps are then carried out (Fig. 3). The first one is the classification and priority marking based on a number of criteria defined by the AM; this step is done by the Classification and Marking Component (CMC). The second step is the processing of the request based on its priority by the Performance Enhancing-Proxy (PEP), and then its possible redirection to the Middleware entity (i.e. server or gateway)

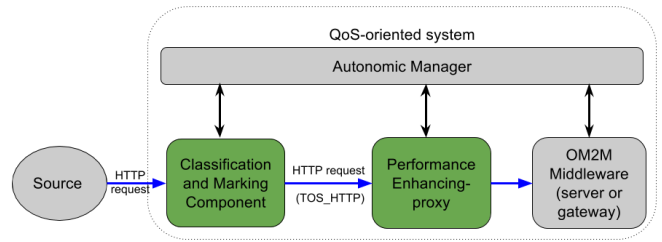

Fig. 3. Architectural composition of request-oriented mechanisms.

#### 1) Classification and Marking Component

The Classification and Marking Component (CMC) adds a TOS_HTTP header (Type Of Service for HTTP) that contains the priority assigned to the HTTP request. The CMC interacts with two main actors:

- **Source:** sender of the HTTP request towards the system. We consider two types of sources: QoS-unaware sources that are not able to express their QoS requirements, and QoS-aware sources that are able to express them through the adding of the TOS_HTTP field with an appropriate priority ;

- **Autonomic Manager:** responsible of the configuration of the underlying components (CMC, PEP and Middleware). For the CMC, the AM intervenes for the classification (requests classes) and the marking (priority of each class) policies.

As described in Fig. 4, the CMC contains the following components:

- **CMC Manager:** acts as a communication interface with the AM; it allows activation/deactivation of the

CMC as well as management of classification and marking policies;

- **CMC Receiver:** receives requests from sources. When activated, it directs the traffic either to the "Request Classifier" if the TOS_HTTP header does not exist, or directly to the "HTTP Forwarder" if it exists or if the component is deactivated;
- **Request Classifier:** classifies the incoming requests according to the classification policy. It first analyses the HTTP request (without TOS_HTTP header) according to the criteria specified in the policy (source IP address, destination IP address, resource URL, etc.), and then redirects it to the "Request Marker";
- **Request Marker:** assigns a priority to each request based on its membership class. It is based on the marking policy and adds the TOS_HTTP header to the HTTP request with the corresponding priority. For instance, the TOS_HTTP can get as a value PRIORITY_HIGH, PRIORITY_MEDIUM or PRIORITY_LOW;
- **CMC Forwarder:** redirects the request containing the TOS_HTTP header to the PEP component if the QoS management is activated, otherwise to the OM2M platform.

- **PEP Controller:** analyses the TOS_HTTP header of the requests and sends them to one of the internal components according to the management policy;
- **PEP Rejecter:** is responsible for rejecting requests based on their priorities. Each priority corresponds to a percentage of rejection;
- **PEP Delayer:** delays requests according to their priorities. Each priority corresponds to a delaying duration;
- **PEP Scheduler:** schedules the incoming requests according to their priorities. This component offers two scheduling policies: priority-first and weighted far queuing;
- **PEP Forwarder:** redirects requests to the OM2M platform for processing.

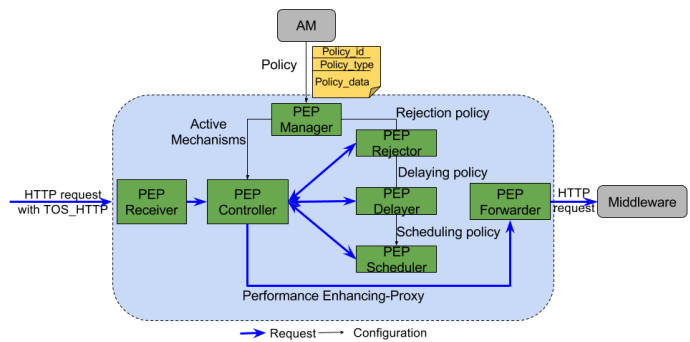

Fig. 5.  Internal composition of the PEP.

The diagram presented by Fig. 6 describes the algorithm implemented by the PEP. M1, M2, and M3 respectively represent the rejecter, delayer and scheduler algorithms.

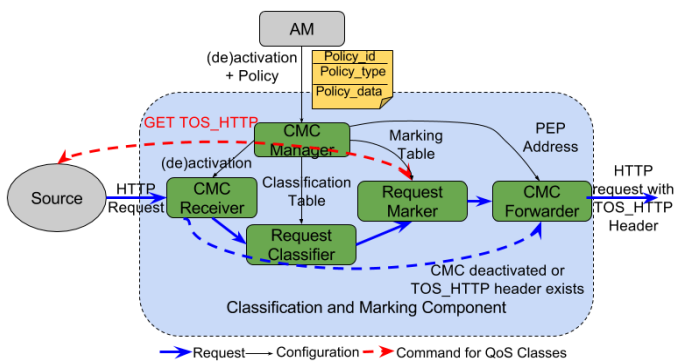

Fig. 4.  Internal composition of the CMC.

*2)  Performance Enhancing Proxy*

The Performance Enhancing Proxy (PEP) is responsible for processing requests according to their priorities (TOS_HTTP). This component can make either request rejection, request delaying, request scheduling, or a combination of these mechanisms depending on the policy communicated by the AM. As illustrated in Fig. 5, the PEP contains the following components:

- **PEP Manager:** applies the policy coming from the AM. It includes the mechanism(s) to be triggered (rejection, delaying or scheduling) as well as the configuration to be applied (percentage of rejection, delaying duration, scheduling policy to be applied and weight of each priority, etc.);
- **PEP Receiver:** receives requests containing the TOS_HTTP header and sends them to the "PEP Controller";

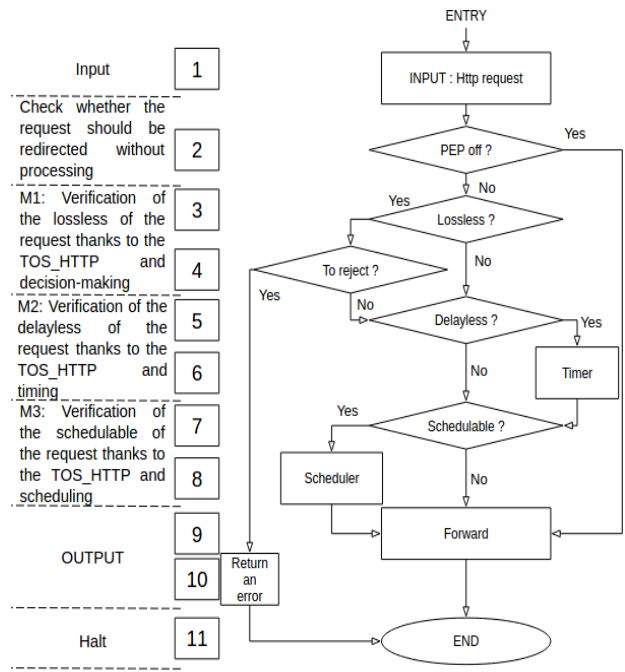

Fig. 6.  PEP algorithm flow chart.

In addition to request-oriented mechanisms, here after we propose resource-oriented mechanisms in order to manage the resources of M2M platforms deployed in dynamic environments such as cloud computing.

### B. Resource-oriented Mechanisms

These mechanisms are inspired by those used in virtualisation deployment environments such as cloud computing. They target the adaptation of the underlying computing resources available in order to improve performances of the Middleware entities. This type of mechanism is more suitable for the server entity that can be run on a cloud platform.

#### 1) Horizontal Scalability oriented Mechanisms

Horizontal scalability consists in adding new instances of the server entity (or part of its components). "Federation" and "Clustering" are two deployment approaches that allow exploitation of the distribution properties of the OM2M platform. Several mechanisms can be used in each approach.

The federation-oriented management that is proposed here consists in the subdivision of the OM2M platform into several components and the dynamic distribution of some of them onto independent virtual machines (VMs). This action can involve, for instance, the Database (DB) by deploying it in another VM (Fig. 7) in order to allow the platform to manage several concurrent accesses and thus reduce the consequent losses.

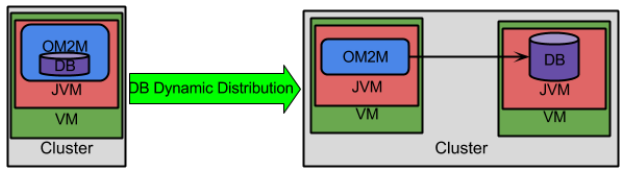

Fig. 7. Distribution mechanism.

Clustering-oriented management mechanisms consist in the creation of several instances of OM2M and the use of a load balancer (Fig. 8) to distribute the HTTP requests. The distribution policy can be guided by several techniques, among them:

- **Round Robin:** fair distribution of the load among all OM2M instances;
- **Weighted Round Robin:** distribution according to weights attributed to each OM2M instance;
- **Load-oriented:** forwarding of the requests to the OM2M instances depending on their load.

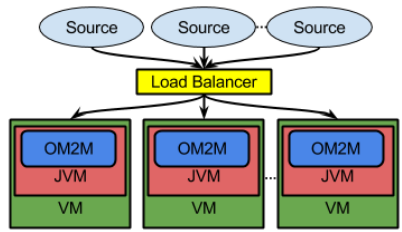

Fig. 8. Load balancing-based adaptation mechanism.

#### 2) Vertical Scalability oriented Mechanisms

Vertical scalability allows adaptation (increasing or decreasing) of the OM2M platform resources in order to support the traffic load. Given that the implementation of the platform is based on the JAVA programming language, three adaptation levels are considered (Fig. 9):

- **dynamic adaptation of the VM resources:** the adapter allows dynamic adjustment (addition or deletion) of VM resources, based on the use of physical resources (processors, memory, physical disk, etc.);
- **dynamic adaptation of JVM resources:** by dynamically taking into account the new resources allocated to the VM;
- **dynamic adaptation of OM2M resources:** OM2M has a number of internal resources (such as threads). The adapter allows dynamic configuration of these resources to improve the performance of OM2M. For example: dynamic configuration of maxThreads to process new requests according to the load.

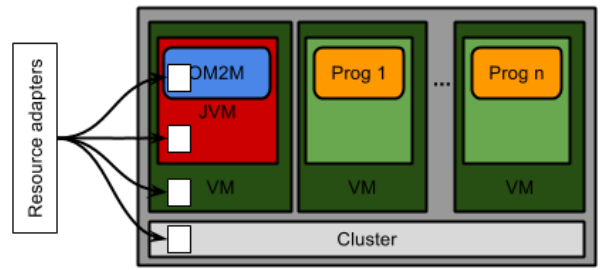

Fig. 9. Multi-level resources adaptation.

In the next section, we give a scenario of application of a request-oriented mechanism in order to demonstrate the benefits in the case of a critical IoT application. Due to space limits, we do not provide a similar scenario demonstrating the benefits of a resource-oriented mechanism.

### III. VALIDATION SCENARIO

This section shows the benefits of the previous mechanisms for a critical application (i.e. having some QoS requirements) through a proof-of-concept. Due to space limits, we only focus on request-oriented mechanisms through the implementation of a rejection policy. Validation is done through performance measurements performed upon an emulation platform allowing generating HTTP requests from different sources towards a real M2M Middleware platform (in our case, the open source OM2M platform).

#### A. Considered ELE and its applications

Let us consider an ELE consisting in a connected nursing home in which the residents are submitted to different kinds of monitoring according to their individual health situation. These monitoring can range from the simple location of patients or their daily diet to potentially very fine supervision, for example during a convalescence period after medical treatment or surgical procedure. Numerous sensors have to be used in the context of multiple applications. In this scenario, we consider three kinds of monitoring applications having characteristics and QoS requirements exposed in Table I.

TABLE I. FEATURES AND QOS REQUIREMENTS OF THE CONSIDERED MONITORING APPLICATIONS

| Application | Average requests rate | Acceptable RTT | Acceptable Loss rate |
|---|---|---|---|
| Postoperative monitoring | 6 req/s | < 350 ms | 0% |
| Location monitoring | 2 req/s | ~10s | 50% |
| Calories monitoring | 2 req/s | No constraint | No constraint |

Let now study how those features and requirements may be taken into account thanks to the proposed QoS oriented mechanism.

### B. The validation platform

As shown in Fig. 10, the validation platform is composed of the following entities:

- **Traffic Emulator:** allows generating http traffic from different IoT sources in order to implement the application scenario. This emulator is based on a controller for injectors configuration. Each injector is able to simulate the traffic of a given application independently of the other injectors. The traffic can be stochastic, periodic or burst;

- **Request-oriented QoS Manager:** consists in the CMC and PEP components. The CMC relies on the IP source address to assign priority and the PEP uses the rejection mechanism;

- **GSCL:** OM2M platform following the SmartM2M standard deployed on a Gateway.

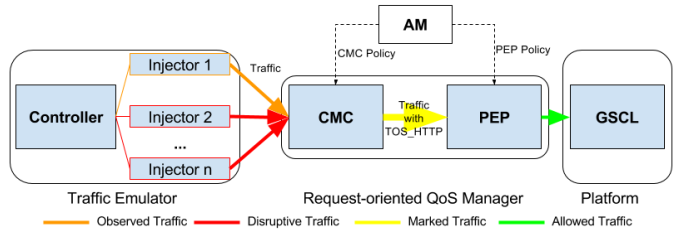

Fig. 10. Validation platform based emulation.

The different components have the physical resources described in Table II. The frequency of each processor is 1.6GHz.

TABLE II. PHYSICAL RESOURCES ALLOCATION

| Resource | Emulator | CMC | PEP | GSCL |
|---|---|---|---|---|
| RAM (MB) | 1024 | 1024 | 2048 | 512 |
| CPU (core) | 1 | 1 | 2 | 1 |

In this scenario, the collected metrics are: RTT (round trip delay), CPU and RAM consumption.

### C. Implementation of the rejection-oriented policy

In this scenario, we consider the monitoring applications introduced in section A. The traffic generated by each application is emulated by a traffic injector. The injectors send HTTP traffic (POST requests) to the Gateway environing the patients. We focus on the injector representing the Postoperative monitoring (PostOp_Inj) for which the system has to respect strong QoS constraints in presence of the other monitoring application flows (Loc_Inj and Food_Inj).

Given those constraints, the CMC assigns a priority to each request coming from a given injector, with respect to its loss tolerance (CMC is guided by the injector ID). It assigns priority HIGH to PostOp_Inj, MEDIUM to Loc_Inj and LOW to Food_Inj. The PEP implements the actions described in Table III depending on the priorities (medium and low priorities (MED_PR & LOW_PR)) in order to meet RTT required by PostOp_Inj. The RTT state is supposed to be triggered when five successive RTT values satisfy the state condition.

TABLE III. ADAPTATION ACTIONS BASED ON THE RTT STATE

| RTT (ms) | State | MED_PR Rejection | LOW_PR Rejection |
|---|---|---|---|
| RTT < 300 | Normal | 0 | 0 |
| 300 ≤ RTT < 400 | Warning | 30 | 70 |
| RTT ≥ 400 | Critical | 40 | 80 |

For adaptation actions, when the PostOp_Inj RTT approaches the required RTT threshold, an alarm is set up in order to anticipate the degradation of the response time.

### D. Results and analysis

Fig. 11 represents the RTT evolution of PostOp_Inj requests in presence of traffic from Loc_Inj and Food_Inj, without QoS mechanisms. We note that the RTT threshold is fast exceeded at the 36[th] request (RTT = 536 ms) and evolves quickly to reach huge RTT values (RTT > 6000 ms). We can then conclude that by default, without QoS mechanisms integration, the system is not able to guarantee a systematic respect of QoS constraints for critical applications.

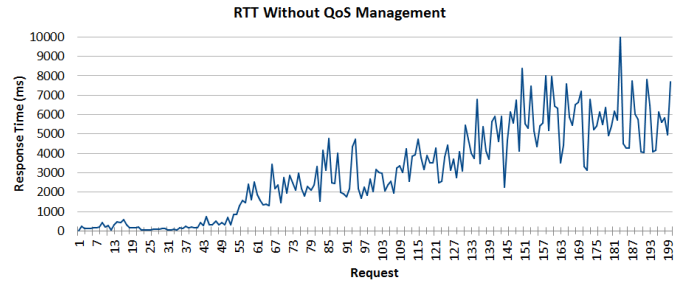

Fig. 11. RTT evolution without QoS management.

By integrating QoS management through CMC and PEP, and applying the rules and actions described above, the RTT

of PostOp_Inj evolves as described by Fig. 12. In this rejection scenario, the RTT of initial requests is less than the "Warning" threshold. After saturation of the gateway, the RTT increases to exceed a threshold of 400ms, which generates a "Critical" state (41th request). The generation of this symptom leads to activate the critical policy allowing the PEP to reject 40% of the Loc_Inj (MEDIUM priority) traffic and 80% of the Food_Inj (LOW priority) traffic. This action reduces the load of the Gateway and allows again having a RTT in a normal state (less than 300ms) from the 57th request.

When this state lasts a certain time (after 12 events), it leads to the generation of a state "Normal" which disables the blocking of requests from other injectors. By applying this policy, the system returns to the initial conditions at the beginning of the scenario (197th request).

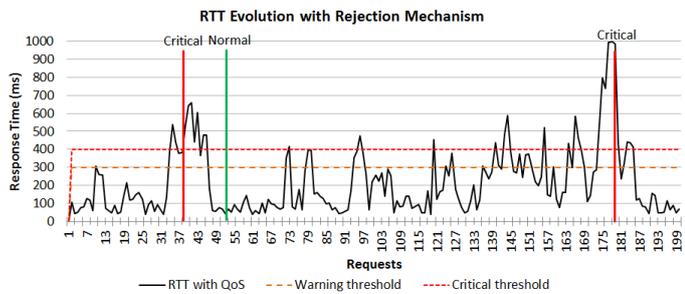

Fig. 12. RTT evolution with rejection mechanism.

In terms of benefices, the average PostOp_Inj RTT for the entire scenario is 222 ms and therefore remains under the threshold (350 ms). Regarding the integration cost of this management, the additional average processing time induced by the CMC and PEP components is 7,992 ms. In terms of resources consumption, the consumed CPU (Fig. 13) of the two mechanisms remains low with an average consumption of 30%.

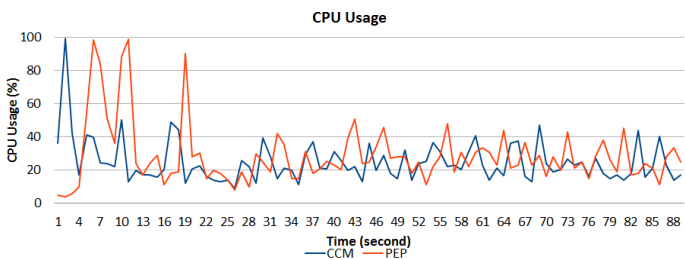

Fig. 13. CPU usage by the CMC and PEP.

In the same time, the RAM consumption (Fig. 14) of the CMC and the PEP increases gradually until becoming insufficient; this is due to the Garbage Collector default configuration.

Based on this validation scenario, the integration of a rejection mechanism provides response times that match the requirements of the Postoperative monitoring application, leading to the improvement of the patient health monitoring and intervention in case of emergency. We can therefore conclude that an IoT system "alone" may not be sufficient, and that it is necessary to integrate a QoS management taking into account constraints of the different applications, in order to enhance the living environments of people (here the patients).

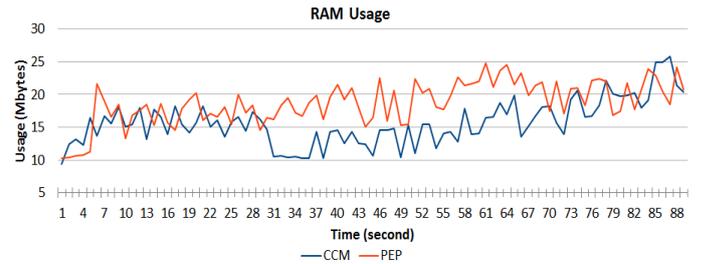

Fig. 14. RAM usage by the CMC and PEP.

## IV. CONCLUSIONS ET PERSPECTIVES

The IoT paradigm is expected to bring ubiquitous intelligence through new applications in order to enhance living and other environments. IoT applications present different features and QoS requirements. Both Network and Middleware underlying levels have to be considered to tackle those QoS requirement. This paper focused on the Middleware level for which QoS-oriented mechanisms are proposed, specified and implemented. Those mechanisms are evaluated through an emulation platform combining applicative traffic emulation and real ETSI M2M compliant Middleware platform. The obtained results allows showing both the need in QoS management and the benefits that can be performed within an instance of ELE thanks to the application of simple QoS-oriented mechanisms.

Our vision is to propose an autonomic QoS-oriented Middleware for the IoT. The current work presents a set of QoS-oriented mechanisms that can be used by the autonomic manager to manage the Middleware platform. The next step will be to integrate the required intelligence to guide the planning component of the Autonomic Manager and propose the most efficient plan. This intelligence can be based on artificial intelligence paradigm in order to choose the adequate mechanism(s) to activate and compute their settings (policies) depending on the characteristics (entity type, deployment context) and the state of the managed entity.